\documentclass[twocolumn,floatfix,prl,showkeys]{revtex4}%
\usepackage{graphicx}%
\usepackage{amsmath}%
\usepackage{amsfonts}
\usepackage{amssymb}
\usepackage{bm}
\usepackage{color}

\def\s{{\sigma}}
\def\e{{\epsilon}}
\def\k{{ {\bm k} }}

\def\q{{ {\bm q} }}
\def\Q{{ {\bm Q} }}

\def\0{{ {\bm 0} }}
\def\w{{\omega}}
\def\a{{\alpha}}
\def\b{{\beta}}

\allowdisplaybreaks[4]

\begin{document}
\title{
Plain $s$-wave superconductivity
near magnetic criticality:
Enhancement of attractive
electron-boson coupling vertex corrections
}
\author{
Rina Tazai$^1$, 
Youichi Yamakawa$^1$, 
Masahisa Tsuchiizu$^2$, and
Hiroshi Kontani$^1$
}


\date{\today }

\begin{abstract}

Recent experiments revealed that the
plain $s$-wave state without any sign-reversal emerges 
in various metals near magnetic criticality.
To understand this counter-intuitive phenomenon,
we study the gap equation for the multiorbital Hubbard-Holstein model,
by analyzing the vertex correction (VC) due to 
the higher-order electron-correlation effects.
We find that the phonon-mediated orbital fluctuations
are magnified by the VC for the susceptibility ($\chi$-VC).
In addition, the charge-channel attractive interaction
is enlarged by the VC for the coupling-constant ($U$-VC),
which is significant when the interaction
has prominent $\q$-dependences; therefore the Migdal theorem fails.
Due to both $\chi$-VC and $U$-VC, the
plain $s$-wave state is caused by the small electron-phonon interaction
near the magnetic criticality against the repulsive Coulomb interaction.
We find that the direct Coulomb repulsion for the plain $s$-wave 
Cooper pair is strongly reduced by the ``multiorbital screening effect.''

\end{abstract}

\address{
$^1$ Department of Physics, Nagoya University,
Furo-cho, Nagoya 464-8602, Japan. 
\\
$^2$ Department of Physics, Nara Women's University, 
Nara 630-8506, Japan
}
 

\sloppy

\maketitle


It is widely believed that the spin-fluctuations are
harmful for the conventional $s$-wave superconductivity.
However, recent experiments have revealed that the
plain $s$-wave state without any sign-reversal emerges in some 
strongly-correlated metals near the magnetic instability.
For example, plain $s$-wave superconductivity with high $T_{\rm c}$ 
is reported in heavily electron-doped FeSe families ($T_{\rm c}=60\sim100$K)
 \cite{Feng-eFeSe-swave,Feng-eFeSe-swave2}
and in A$_3$C$_{60}$ (A =K, Rb, Cs; $T_{\rm c}>30$K)
 \cite{A3C60}.
In both compounds, electron-phonon ($e$-ph) interaction 
may play a crucial role in the pairing mechanism, as discussed in 
Refs. \cite{Shen-replica,Millis,Choi,Jhonston,Capone,A3C60-ph-theory1,Kim}.
Even so, a fundamental question is why 
the high-$T_{\rm c}$ plain $s$-wave state appears against the 
repulsive interaction by spin-fluctuations.
More surprisingly, the plain $s$-wave state is reported
in heavy-fermion superconductor CeCu$_2$Si$_2$ near the magnetic phase, 
according to the measurements of the specific heat, penetration depth, 
thermal conductivity, and electron irradiation effect on $T_{\rm c}$
 \cite{Matsuda-CeCuSi,Kittaka-CeCuSi}.

Therefore, it is a significant problem for theorists to 
establish a general mechanism of the plain $s$-wave superconductivity
in strongly correlated electron systems.
One important feature of these $s$-wave superconductors
would be the orbital degrees of freedom.
In this case, in principle, the pairing glue for the plain $s$-wave state 
may be realized by the orbital fluctuations.
The two possible origins of the orbital fluctuations are 
the higher-order many-body process given by the vertex correction (VC)
\cite{Onari-SCVC} 
and the $e$-ph interaction
\cite{Kontani-RPA}.
The significant questions are (i) whether these
two different origins of the orbital fluctuations
({\it i.e.}, the VC due to Coulomb interaction and the $e$-ph interaction)
cooperate or not, 
(ii) why the high-$T_{\rm c}$ plain $s$-wave state is realized 
against the strong magnetic fluctuations, and 
(iii) how the plain $s$-wave Cooper pairs escape from the 
strong direct Coulomb repulsion in multiorbital systems.

In this paper, 
we analyze a canonical two-orbital model in detail
in order to resolve the above-mentioned fundamental questions (i)-(iii).
We study the pairing mechanism
in the presence of strong magnetic fluctuations and
small phonon-mediated attractive interaction,
by considering the VC for the orbital susceptibility ($\chi$-VC)
and the VC for the pairing interaction ($U$-VC) consistently.
In both VCs, the significant contributions come from the  
Aslamazov-Larkin (AL) processes,
which represent the strong orbital-spin interference 
driven by the electron correlation
\cite{Onari-SCVC}.
Due to both VCs, weak $e$-ph interaction is enough to realize 
the high-$T_{\rm c}$ plain $s$-wave state near the magnetic quantum criticality.
The present theory explains
the characteristic phase diagram in 
typical strongly-correlated plain $s$-wave superconductors
such as FeSe, A$_3$C$_{60}$, and CeCu$_2$Si$_2$.
We also find that the direct Coulomb repulsion for the intra-orbital 
$s$-wave Cooper pair is reduced by the ``multiorbital screening effect.''


\begin{figure}[!htb]
\includegraphics[width=.85\linewidth]{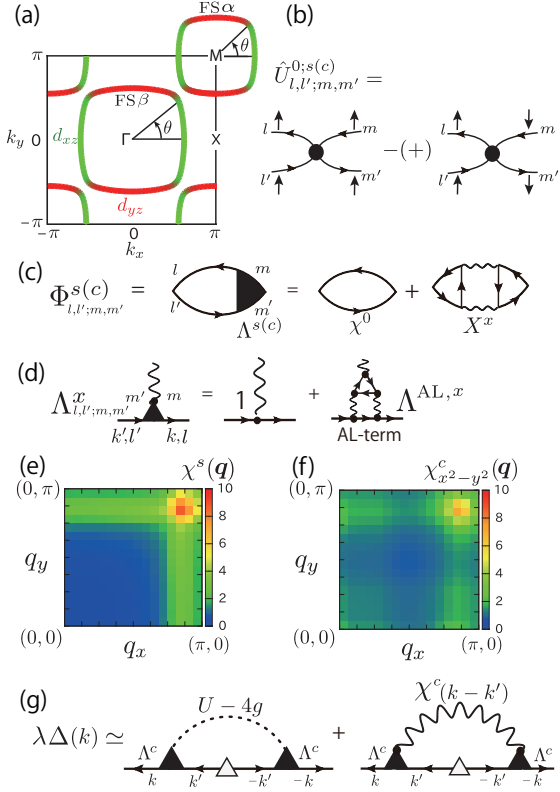}
\caption{
(Color online)
(a) FSs of the two-orbital model
composed of $d_{xz}$ (green) and $d_{yz}$ (red) orbitals.
The nesting between FS $\alpha$ and FS $\beta$ 
causes the spin and orbital fluctuations at $\Q\simeq (0.8\pi,0.8\pi)$. 
(b) Multiorbital Coulomb interaction for the spin- (charge-) channel.
(c) Irreducible susceptibility with the $\chi$-VC.
(d) Dressed electron-boson coupling due to the AL process.
The $\q$-dependences of the 
(e) total spin susceptibility $\chi^{s}(\q)$ and
(f) orbital susceptibility $\chi^{c}_{x^2-y^2}(\q)$ at
$(\a_S,\a_C)=(0.92,0.93)$.
(g) Linearized gap equation with the three-point VC ($U$-VC).
}
\label{fig:fig1}
\end{figure}

We start from the two-orbital Hubbard-Holstein model
on the square lattice $H=H_{0}+H_{U}+H_{\rm ph}$, where
$H_{0}=\sum_{\k,\sigma}\sum_{l,m}\xi_{\k}^{l,m}
c^{\dagger}_{\k l\sigma}c_{\k m\sigma}$ is the kinetic term, and
$H_{U}=\frac{1}{4}\sum_{i}\sum_{l l' m m'}
\sum_{\sigma \sigma' \rho \rho'}
U_{l,l';m,m'}^{0;\sigma \sigma' \rho \rho'}
c^{\dagger}_{i l \sigma}
c_{i l' \sigma'}
c_{i m \rho} 
c^{\dagger}_{i m' \rho'}$
is the on-site Coulomb interaction term.
Here, $i$ is a lattice site index, $c^{\dagger}_{\k l \sigma}$ ($c_{\k l \sigma}$) is the creation 
(annihilation) operator of $d$-electrons with wave-vector $\k$,
orbital $l$, and spin $\sigma$.
$l=1$ ($2$) represents the orbital $d_{xz}$ ($d_{yz}$). 
$\xi^{l,m}_{\k}$ is defined as 
$\xi^{1,1}_{\k}=-2t\cos k_{x} -2t^{''}\cos k_{y}$, 
$\xi^{2,2}_{\k}=-2t\cos k_{y} -2t^{''}\cos k_{x}$, and
$\xi^{1,2}_{\k}=\xi^{2,1}_{\k}=-4t^{'}\sin k_{x} \sin k_{y}$.
Hereafter, we set the hopping parameters as
($t^{}$, $ t^{'}$, $ t^{''})=(1, 0.1, 0.1)$.
The unit of energy in the present study is $t=1$, and 
the electron filling is fixed as $n_{e}=2.30$.
The two Fermi surfaces (FSs), FS $\alpha$ and FS $\beta$,
are shown in Fig. \ \ref{fig:fig1}(a), where $\theta$ is
the angle of the $\k$ on each FS.
The bare multiorbital Coulomb interaction
 $U_{l,l';m,m'}^{0;\sigma \sigma' \rho \rho'}$
is composed of the intra-orbital Coulomb interaction $U$,
inter-orbital one $U'$, Hund's coupling $J$, and pair hopping $J'$
\cite{Takimoto}.
$U_{l,l';m,m'}^{0;\sigma \sigma' \rho \rho'}$ is 
uniquely decomposed into the spin-channel and charge-channel;
${\hat U}^{0;\sigma \sigma' \rho \rho'}
=\frac{1}{2}{\hat U}^{0;s}
\vec{\bf{\sigma}}_{\sigma \sigma'} \cdot \vec{\bf{\sigma}}_{\rho' \rho}
+\frac{1}{2}{\hat U}^{0;c}\delta_{\sigma,\sigma'}\delta_{\rho',\rho}$,
%
%
where $\vec{\bf{\sigma}}$ is the Pauli matrix vector and $\hat{U}^{0;s(c)}$ is  
the spin- (charge-) channel Coulomb interaction
shown in Fig. \ref{fig:fig1}(b). Their expressions are
given in the Supplementary Material (SM):A \cite{SM}.
Hereafter, we simply put $J=J'=(U-U')/2$.

In addition, $H_{\rm ph}$ is the phonon-related term given by
$H_{\rm ph}=\omega_{D}\sum_{i}
b^{\dagger}_{i}
b_{i}+\eta \sum_{i}
(b^{\dagger}_{i}+b_{i})
(\hat{n}_{i}^{xz}-\hat{n}_{i}^{yz}) 
$,
where $\hat{n}_{i}^l$ is an electron number operator for orbital $l$,
$b^{\dagger}_{i}$ ($b_{i}$) is a phonon creation 
(annihilation) operator,
$\eta$ is the coupling constant between electrons and 
$B_{1g}$-symmetry phonon, and
$\omega_{D}$ is the phonon frequency.
The phonon-mediated retarded interaction is
$-g(\omega_{j})\sum_{i}(\hat{n}_{i}^{xz}-\hat{n}_{i}^{yz})(\hat{n}_{i}^{xz}-\hat{n}_{i}^{yz})$
\cite{Kontani-RPA},
%
where
$g(\omega_{j})=g\frac{\omega_{D}^2}{\omega_{D}^{2}+\omega_{j}^{2}}$
and 
$g=\frac{2\eta^{2}}{\omega_{D}}\ (>0)$.
$\omega_{j}=2j\pi T$ is the boson Matsubara frequency with integer $j$.
In the present model, $B_{1g}$ orbital fluctuations are induced by the
$\chi$-VC even for $g=0$. 
Due to the $\chi$-VC, the $B_{1g}$ orbital
fluctuations are strongly enhanced by introducing the small $g$
given by the $B_{1g}$ phonon.
Such enhancement is not realized by non $B_{1g}$ phonons.


Now, we derive the spin and charge susceptibilities
by analyzing the $\chi$-VC for the charge-channel self-consistently,
based on the self-consistent vertex correction (SC-VC) method \cite{Onari-SCVC}. 
Hereafter, we fix the parameters $J/U=0.08$ and $T=5\times 10^{-2}$, 
and use the notations $k=(\k,\e_n)=(\k,(2n+1)\pi T)$ and
$q=(\q,\w_j)=(\q,2j\pi T)$.
We adopt $N_{k}= 32\times 32$ $\k$-meshes 
and $256$ Matsubara frequencies.
In the present model, the spin ($x=s$) and charge ($x=c$) susceptibilities
are
\begin{eqnarray}
\hat{\chi}^{x}(q)= {\hat \Phi}^x(q)
[{\hat 1}-{\hat C}^x{\hat \Phi}^x(q)]^{-1} ,
 \label{eqn:chi}
\end{eqnarray}
%
%
where $\hat{\chi}^{x}$, $\hat{\Phi}^{x}$, and $\hat{C}^{x}$ 
are $2^{2}\times 2^{2}$ matrices in the orbital basis,
 and ${\hat \Phi}^x(q)=\hat{\chi}^{0}(q)+\hat{X}^{x}(q)$
is the irreducible susceptibility. 
We explain the matrix expressions of $\hat{\chi}^{x}(q)$ and
$\hat{X}^{x}(q)$ in the SM:B and SM:C, respectively \cite{SM}.
$\hat{X}^{x}(q)$ is the $\chi$-VC given by the AL process.
Its diagrammatic expression is shown in Fig. \ref{fig:fig1}(c),
which contains the three-point VC, $\hat{\Lambda}^x(k,k')$,
as shown in Fig. \ref{fig:fig1}(d).
The solid and wavy lines represent the electron Green 
function ${\hat G}(k)$
and ${\hat \chi}^{x}(q)$, respectively.
The bare susceptibility is given by
$\chi_{l,l';m,m'}^0(q)= -\frac{T}{N_k}
\sum_{k}G_{l,m}(k+q)G_{m',l'}(k)$,
where $G_{l,m}(k)$ is the Green function in the orbital basis without the self-energy.
The spin (charge) Stoner factor $\a_{S(C)}$ is given by 
the largest eigenvalue of ${\hat C}^{s(c)}{\hat \Phi}^{s(c)}(q)$.
Here, we calculate $\hat{X}^{c}(q)$ self-consistently,
by neglecting $\hat{X}^{s}(q)$ since it is less important 
\cite{Yamakawa-FeSe}. 
In the random phase approximation (RPA), both $\hat{X}^{s}$ and $\hat{X}^{c}$ are dropped.

As we explain in the SM:B \cite{SM}, the interaction terms are ${\hat C}^s\equiv\hat{U}^{0;s}$ 
and ${\hat C}^c\equiv\hat{U}^{0;c}-\hat{g}(\omega_{j})$, where
$\hat{g}$ is the phonon-mediated interaction given as
$ g_{l,l';m,m'}(\omega_{j})\equiv-2g(\omega_{j})\cdot \delta_{l,l'}\delta_{m,m'}
(2\delta_{l,m}-1)$.
Here, we neglect the ladder-diagram for the phonon-mediated interaction 
by assuming the relation $\omega_{D} \ll W_{\rm band}$ (bandwidth).
In this case, $\a_S$ is independent of $g$ if we put $\hat{X}^{s}=0$  
 \cite{Kontani-RPA}.
%
%
Note that the orbital-fluctuations are caused by the 
cooperation between $\chi$-VC and $g$,
even when $\omega_{D} \gg W_{\rm band}$.

Figures \ref{fig:fig1}(e) and \ref{fig:fig1}(f) show the
obtained total spin susceptibility,
$\chi^{s}({\bm q})=\sum_{l , m}\chi^{s}_{l,l;m,m}({\bm q})$, and
orbital susceptibility with respect to the 
$B_{1g}$ orbital operator $\hat{n}_{xz}-\hat{n}_{yz}$,
$\chi^c_{x^2-y^2}({\bm q})=\sum_{l , m}(-1)^{l+m}\chi^{c}_{l,l;m,m}({\bm q})$,
respectively. 
The used parameters are $(U,g)=(2.1,0.15)$, and
the realized Stoner factors are
$(\a_S,\a_C)=(0.92,0.93)$.
The spin susceptibility is enlarged due to the nesting between FS $\a$ and FS $\b$, and
the strong orbital susceptibility is induced by the $\chi$-VC and the small $g$ 
is due to the $B_{1g}$ phonon.
The antiferro-orbital ordered phase is realized when $\a_{C}\geq1$.
More detailed results are shown in the SM:B \cite{SM}.



Next, we analyze the linearized gap equation beyond the 
Migdal-Eliashberg scheme given as \cite{Yamakawa-eFeSe-SC}
\begin{eqnarray}
\lambda \Delta^{a}(\theta,\epsilon_{n})=
-\frac{\pi T}{(2\pi)^2}\sum_{a',\epsilon_{m}}
\int_{0}^{2\pi} \frac{d\theta'}{v_{a',\theta'}}\left| \frac{\partial \k_{a',\theta'}}{\partial \theta'} \right|
\nonumber \\ 
\times \frac{\Delta^{a'}( \theta',\epsilon_{m})}{|\epsilon_{m}|} V^{a, a'}(\theta,\epsilon_n,\theta',\epsilon_m) ,
\label{eqn:linear}
\end{eqnarray}
which is diagrammatically expressed in Fig.\ \ref{fig:fig1}(g).
$\Delta^{a}(\theta,\epsilon_{n})$ and $\lambda$ are the 
singlet superconducting gap function on the FS $a$ ($a=\a,\b$)
and its eigenvalue, respectively. 
${V}^{a,a'}(\theta,\epsilon_n,\theta',\epsilon_m)$ is the 
pairing interaction in the band basis.
$\k_{a,\theta}$ and $v_{a,\theta}$ are the Fermi momentum 
and the Fermi velocity on FS $a$, respectively.

Using $\hat{\chi}^{s(c)}(q)$ derived from the SC-VC method,
the paring interaction in the orbital basis is given as
 \cite{Yamakawa-eFeSe-SC}
\begin{eqnarray}
\hat{V}(k,k')=
\frac{3}{2}{\hat I}^{\Lambda,s}(k,k')-
\frac{1}{2}{\hat I}^{\Lambda,c}(k,k')-
\hat{C}^{s},
\label{eqn:V}
\end{eqnarray}
which is transformed to $V^{a,a'}(\theta,\e_{n},\theta',\e_{m})$
by using the unitary matrix $u_{l,a}(\k)=\langle l,\k|a,\k \rangle$.
Here,
${\hat I}^{\Lambda,x}(k,k')={\hat \Lambda}^{x}(k,k')
{\hat I}^{x}(k-k') {\hat {\bar \Lambda}}^{x}(-k,-k')$,
and 
${\hat I}^{x}(k-k')={\hat C}^{x}{\hat \chi}^{x}(k-k'){\hat C}^{x}+{\hat C}^{x}$.
${\hat \Lambda}^{x}(k,k')$ is the three-point vertex with the 
AL-type $U$-VC shown in Fig. \ref{fig:fig1}(d),
and ${\bar \Lambda}^x_{l,l';m,m'}(k,k')\equiv{\Lambda}^x_{m',m;l',l}(k,k')$
\cite{Yamakawa-eFeSe-SC}.
Then, the effective Coulomb interaction dressed by the $U$-VC is
$\hat{U}^{x}(k,k')={\hat \Lambda}^x(k,k'){\hat C}^{x}$, which is 
given in the SM:C \cite{SM}.
Since the contribution to the $U$-VC from $\chi^s(q)$ 
dominates over the one from $\chi^c(q)$ even for 
$\a_S\sim\a_C$, as explained in Refs. \cite{Yamakawa-eFeSe-SC,Yamakawa-FeSe}, 
we can safely set $g=0$ in\\for calculating ${\hat \Lambda}^x$.

\begin{figure}[!htb]
\includegraphics[width=.85\linewidth]{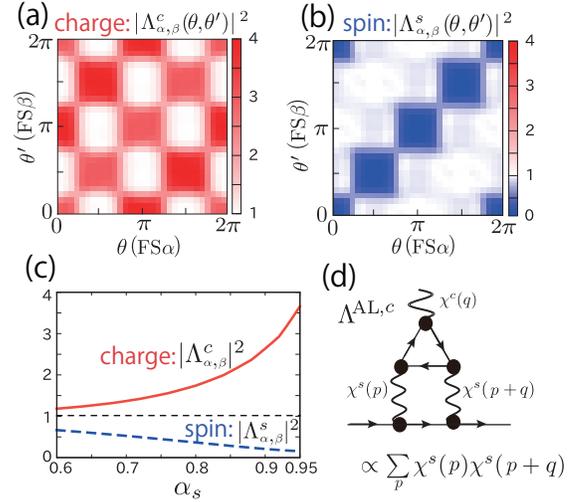}
\caption{
(Color online)
(a) $|\Lambda^{c}_{\a,\b}(\theta,\theta')|^2$ and 
(b) $|\Lambda^{s}_{\a,\b}(\theta,\theta')|^2$ 
for the lowest frequency.
$\theta$ ($\theta')$ represents the Fermi point on the FS $\a$ ($\b$).
(c) $\a_S$ dependence of 
$|\Lambda^{s(c)}_{\a,\b}(\theta,\theta')|^2$ at 
$\theta=\theta'=0$.
(d) AL-type $U$-VC for the charge channel.
}
\label{fig:fig2}
\end{figure}

Here, we explain the important role of the $U$-VC on the superconductivity.
In Figs. \ref{fig:fig2}(a) and \ref{fig:fig2}(b), we show the charge- and spin-channel enhancement factors in the band basis at $\e_{n}=\e_{n'}=\pi T$ defined as
\begin{eqnarray}\left|\Lambda^{s(c)}_{a,a'}(\theta,\theta')\right|^{2} \equiv
\Big|\sum_{l,l',m}
 \Lambda^{s(c)}_{l,l';m,m}(k,k')
u^{*}_{l,a}(\theta)
u_{l',a'}(\theta')\Big| ^{2}.
\end{eqnarray}
Here, we set $U=2.1$ ($\a_S=0.92$).
Figure \ref{fig:fig2}(a) means that $|\Lambda^c|^2\gg1$, 
when both Fermi points ($\theta$ and $\theta'$) are composed of the same orbital.
In contrast, Fig. \ref{fig:fig2}(b) means that 
$|\Lambda^s|^2\ll1$, for the same orbital.
Their $\a_S$-dependences are shown in  Fig. \ref{fig:fig2} (c),
which are very similar to Fig. 8 (a) for $n_{e}=2.67$ in Ref. \cite{Rina}.
The obtained relation $|\Lambda^c|^2\gg1$ for $\a_S\lesssim1$
originates from the AL-type $U$-VC for the charge-channel 
$\Lambda^{{\rm AL},c}(q) \propto \sum_{p} \chi^{s}(p)\chi^{s}(p+q)$, which is
explained in Fig. \ref{fig:fig2} (d) and in Ref. \cite{Yamakawa-eFeSe-SC}.
This relation owing to the AL-processes
has been confirmed by the functional-renormalization-group
(fRG) analysis in Refs. \cite{Rina},
In the fRG method, the higher-order VCs,
even higher order than Figs. \ref{fig:fig1} (c) and (d), 
are generated in a systematic and unbiased way.

Hereafter, we solve the gap equation (\ref{eqn:linear}) numerically
and simply set $g(\omega_{j})=g$ by neglecting the retardation effect.
This approximation leads to the underestimation of 
the plain $s$-wave state.
In addition, we neglect both the $U$-VC and $\chi$-VC for finite $\omega_{j}$,
and also drop the crossing pairing interaction introduced in 
Ref. \cite{Yamakawa-eFeSe-SC}.
These simplifications also lead to the underestimation of the 
plain $s$-wave state
\cite{Yamakawa-eFeSe-SC}.
We summarize the approximations 
used in the numerical study in the SM:C \cite{SM}.

\begin{figure}[!htb]
\includegraphics[width=.85\linewidth]{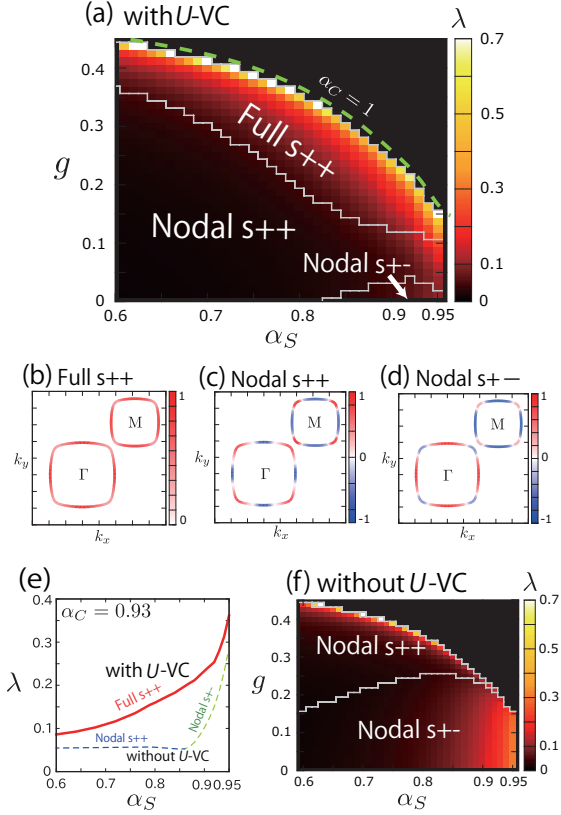}
\caption{(Color online) 
(a) Obtained phase diagram for the $s$-wave states
in the presence of the $U$-VC.
The gap functions for the
(b) full-gap $s_{++}$ state at $g=0.2$,
(c) nodal $s_{++}$ state at $g=0.06$, and
(d) nodal $s_{+-}$ state at $g=0.04$, for $\a_{S}=0.92$.
The antiferro-orbital order occurs
when $\a_{C} \geq 1$.
(e) $\a_S$ dependence of $\lambda$ at $\a_{C}=0.93$.
The eigenstate is full-gap $s_{++}$ state with $U$-VC,
whereas it is nodal $s_{++(+-)}$ state without $U$-VC.
(f) Phase diagram obtained by neglecting the $U$-VC.
}
\label{fig:fig3}
\end{figure}
In Fig. \ref{fig:fig3}(a), we show the largest eigenvalue 
with the phase boundary between three $s$-wave states 
given in Figs. \ref{fig:fig3}(b)-\ref{fig:fig3}(d),
which we call the phase diagram below.
In the nodal $s_{++}$ ($s_{+-}$) state,
$\Delta^\a(N\pi/2,\epsilon_{n})$ and $\Delta^\b(N\pi/2,\epsilon_{n})$ 
have the same (opposite) sign for $N=0,1,2,3$.
In Fig. \ref{fig:fig3}(a) and Fig. S3(a) in the SM:D \cite{SM}, 
the full-gap $s_{++}$ state without any sign reversal
corresponds to the largest eigenvalue for a wide region with $\a_C\geq 0.8$.
The obtained $\lambda$ for the full-gap $s_{++}$ state is very large, since the attractive (repulsive) interaction is enlarged (suppressed) by $|\Lambda^{c(s)}|^2$ \cite{Yamakawa-eFeSe-SC,Rina,Onari-SCVCS}.
Thus, $T_{c}$ for the full-gap $s$-wave state is expected to be high.
In contrast, $\lambda$ for nodal $s_{++}$ and $s_{+-}$ states
is very small. 

Figure \ref{fig:fig3}(e) shows the
$\a_S$-dependence of $\lambda$ at $\a_{C}=0.93$.
When the $U$-VC is included, $\lambda$ for the full-gap $s_{++}$ 
state drastically increases with $\a_{S}$.
In contrast, the full-gap $s_{++}$ state 
disappears in the phase diagram if the $U$-VC is neglected (=Migdal approximation \cite{Mig})
, as shown in Fig. \ref{fig:fig3}(f).
Although the paring interaction has the small energy scale,
the Migdal theorem fails 
due to its strong $\q$-dependence.
Therefore, the significance of the $U$-VC for the plain $s$-wave state
is clearly confirmed.

We see in Fig. \ref{fig:fig3}(a) that, the strong orbital fluctuations
$\a_C\lesssim 1$ are realized just $g\approx 0.15$ $(\lesssim U/10)$ for $\a_S\gtrsim 0.9$. 
By following Ref. \cite{Ohno-SCVC},
orbital susceptibility for $\a_C\lesssim1$ is approximately given as 
$\chi^{c}_{x^2-y^2}(\Q)\sim {\Phi}^{c}(\Q)[{1-(2U'-U+4g){\Phi}^{c}(\Q)}]^{-1}$,
where ${\Phi}^{c}(\Q)$ is the intra-orbital irreducible susceptibility.
Due to the existence of $U'$,
$\a_C=(2U'-U+4g){\Phi}^{c}(\Q)$ reaches unity 
by introducing the small $g$ due to the $B_{1g}$ phonon.
In addition, the required $g$ for $\a_C=1$ is reduced
if the relation ${\Phi}^{c}(\Q)\gg \chi^{0}(\Q)$ is realized 
by the AL-VC.
Therefore, the strong orbital fluctuations are induced 
by the cooperation between the $B_{1g}$ phonon ($g$) and the $\chi$-VC.

To summarize, the full-gap $s_{++}$ wave state is stabilized by the charge-channel pairing interaction 
$V^{c}\simeq \frac{1}{2} \{U-4g+(2U'-U+4g)^{2}\chi^c_{x^2-y^2}(\Q)\}|\Lambda^{c}|^{2}$,
which takes large negative value when $\a_C \lesssim 1$
and $|\Lambda^{c}|^{2}\gg 1$. The latter condition is realized when $\a_{S} \lesssim 1$ 
due to the AL-VC.
We verified in the SM:E \cite{SM} that 
the full-gap $s_{++}$
state corresponds to the largest eigenvalue for a wide filling range
if the $U$-VC is included in the gap equation.
Note that the momentum dependence of the 
$\Lambda^{x}(k,k')$ is quite important
since the full-gap $s_{++}$ phase disappears
if we apply the local approximation to $U$-VC;
$\Lambda^x_{\rm loc}(\e_n,\e_{n'}) \equiv 
\langle \Lambda^x(k,k') \rangle_{\k,\k'\in{\rm FS}}$, as shown in the SM:D \cite{SM}.
In the SM:F and SM:G, we discuss that the full-gap $s_{++}$ wave state is further stabilized by 
introducing the dilute impurities and by considering the retardation effect, respectively.

\begin{figure}[!htb]
\includegraphics[width=.9\linewidth]{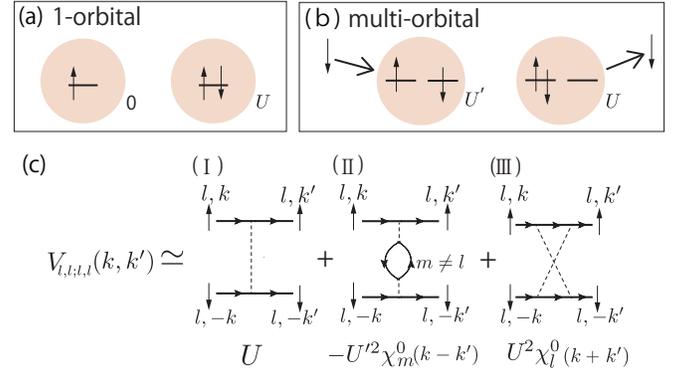}
\caption{
(Color online)
Schematic depairing processes for the intra-orbital Cooper pair
caused by the direct Coulomb interaction,
in (a) one-orbital and (b) two-orbital cases.
In (b), the energy cost for the intra-orbital pair
is reduced to $\sim (U-U')$ due to the ``multiorbital screening effect.''
(c) Pairing interaction for the intra-orbital Cooper pair up to second-order.
The process (II) exists only in multiorbital models. Here, $\chi^{0}_{l}\equiv \chi^{0}_{l,l;l,l}$.
}
\label{fig:fig5}
\end{figure}

Finally, we explain that the direct Coulomb repulsion for the $s$-wave 
Cooper pair is strongly reduced by the ``multiorbital screening effect.''
Figures \ref{fig:fig5}(a) and \ref{fig:fig5}(b) show the schematic depairing processes
for the ``intra-orbital Cooper pair''
for the single- and multi-orbital models, respectively.
Here, the effect of $J$ is neglected for simplicity.
The energy cost for the intra-orbital Cooper pair,
which is $\sim U$ in the single-orbital model, 
is drastically reduced to $\sim (U-U')$ in the multi-orbital model.
This reduction is caused by the screening due to the 
electrons on other orbitals.
Thus, we call this effect the multiorbital screening.
Figure \ref{fig:fig5}(c) shows the pairing interaction
for the intra-orbital Cooper pair up to second-order.
We assume $G_{l,m}=G_{l}\cdot \delta_{l,m}$ for simplicity.
The process (II), which exists only in the multiorbital models,
reduces the direct Coulomb depairing given by (I).
This multiorbital screening effect is prominent when 
$U'\chi^0_{m,m;m,m}\sim O(1)$ for $m\ne l$.
Note that the depairing is suppressed further 
by the retardation effect.

We comment on other theoretical studies.
Based on the dynamical-mean-field-theory (DMFT)
or variational cluster approach (VCA),
mechanisms of the plain $s$-wave state due to the electron correlation
(together with the $e$-ph interaction)
have been discussed in Refs. 
\cite{Capone,A3C60-ph-theory1,Kim,Pruschke,Yamamoto}.
In CeCu$_2$Si$_2$, both the valence fluctuation and 
orbital fluctuation scenarios have been discussed in Refs. \cite{Miyake,Hattori}.

In summary, we proposed the mechanism of the 
plain $s$-wave state in strongly correlated metals
with the weak $B_{1g}$ $e$-ph interaction.
We demonstrated that the strong orbital fluctuations emerge
due to the cooperation between the $\chi$-VC and the $B_{1g}$ $e$-ph interaction,
and the orbital-fluctuation-mediated attractive force is enhanced 
by the charge-channel $U$-VC.
In contrast, the repulsive force due to the spin fluctuations 
is reduced by the spin-channel $U$-VC.
In addition, the direct Coulomb repulsion for the intra-orbital cooper
pair is strongly reduced by the multiorbital screening effect.
The plain $s$-wave state 
has a large eigenvalue in the vicinity of the magnetic quantum
criticality as shown in Fig. \ref{fig:fig3}(e).
The present theory may explain  
the strongly-correlated plain $s$-wave superconductivity
in FeSe, A$_3$C$_{60}$, and CeCu$_2$Si$_2$.


We stress that the charge-channel $U$-VC 
is enhanced by the AL-VC even in one-orbital models, which 
explains the result of the quantum Monte Carlo study
for the two-dimensional one-orbital Hubbard model in Ref. \cite{Scalapino}.

\acknowledgments
We are grateful to Y. Matsuda, T. Shibauchi, S. Kasahara, and S. Onari
for their useful comments and discussions.
This study has been supported by Grants-in-Aid for Scientific
Research from MEXT of Japan.


\clearpage

\makeatletter
\renewcommand{\thefigure}{S\arabic{figure}}
\renewcommand{\theequation}{S\arabic{equation}}
\makeatother
\setcounter{figure}{0}
\setcounter{equation}{0}
\setcounter{page}{1}
\setcounter{section}{1}

\begin{widetext}
\begin{center}
{\Large [Supplementary Material]}
\end{center} 

\begin{center}
{\large
\textbf{
Plain $s$-wave superconductivity
near the magnetic criticality:
Enhancement of attractive electron-boson coupling vertex corrections
}}
\end{center} 
\begin{center}
Rina Tazai,
Youichi Yamakawa, 
Masahisa Tsuchiizu, and
Hiroshi Kontani
\end{center} 

 
\end{widetext}

\subsection{A: Multiorbital Coulomb interaction}

In the main text, we studied the two-orbital Hubbard-Holstein model.
Hereafter, we use the variables $a\sim h,l,l',m,m'$ as 
orbital indices in this Supplementary Material (SM).
First, we explain the multiorbital Coulomb interaction,
which is uniquely decomposed into 
the spin-channel and charge-channel parts as
\cite{SYamakawa-eFeSe-SC}
\begin{equation}
 U_{l,l';m,m'}^{0;\sigma \sigma' \rho \rho'}
=\frac{1}{2}U^{0;s}_{l,l';m,m'}
\vec{\bf{\sigma}}_{\sigma \sigma'} \cdot \vec{\bf{\sigma}}_{\rho' \rho}
+\frac{1}{2}U^{0;c}_{l,l';m,m'}\delta_{\sigma,\sigma'}\delta_{\rho',\rho} .
\end{equation}
Here, the matrix elements of the spin- and charge-channel 
Coulomb interactions are
\begin{eqnarray} U^{0;s}_{l,l';m,m'}&=&\begin{cases}
U &(l=l'=m=m')\\
U' &(l=m\neq l'=m')\\
J &(l=l'\neq m=m')\\
J' &(l=m'\neq l'=m)\\
0 &(\rm{otherwise})\end{cases},\\
U^{0;c}_{l,l';m,m'}&=&
\begin{cases}
-U &(l=l'=m=m')\\
U'-2J &(l=m\neq l'=m')\\
-2U'+J &(l=l'\neq m=m')\\
-J' &(l=m'\neq l'=m)\\
0 &(\rm{otherwise})\end{cases}.
\end{eqnarray}
%
%
\subsection{B: Multiorbital spin and charge susceptibilities}

In the main text, we use 
the $2^2\times2^2$ matrix representation for
the multiorbital spin (charge) susceptibility
$\hat{\chi}^{x}(q) (x=s,c)$.
The matrix elements of  $\hat{\chi}^{x}(q)$ in Eq. (\ref{eqn:chi}) in the main text
is given as $\chi^{x}_{l,l';m,m'}(q)=\sum_{a,b} \Phi^{x}_{l,l';a,b}(q)
[\hat{1}-\hat{C}^{x}\hat{\Phi}^{x}(q)]^{-1}_{a,b;m,m'}$.

\begin{figure}[!htb]
\includegraphics[width=.7\linewidth]{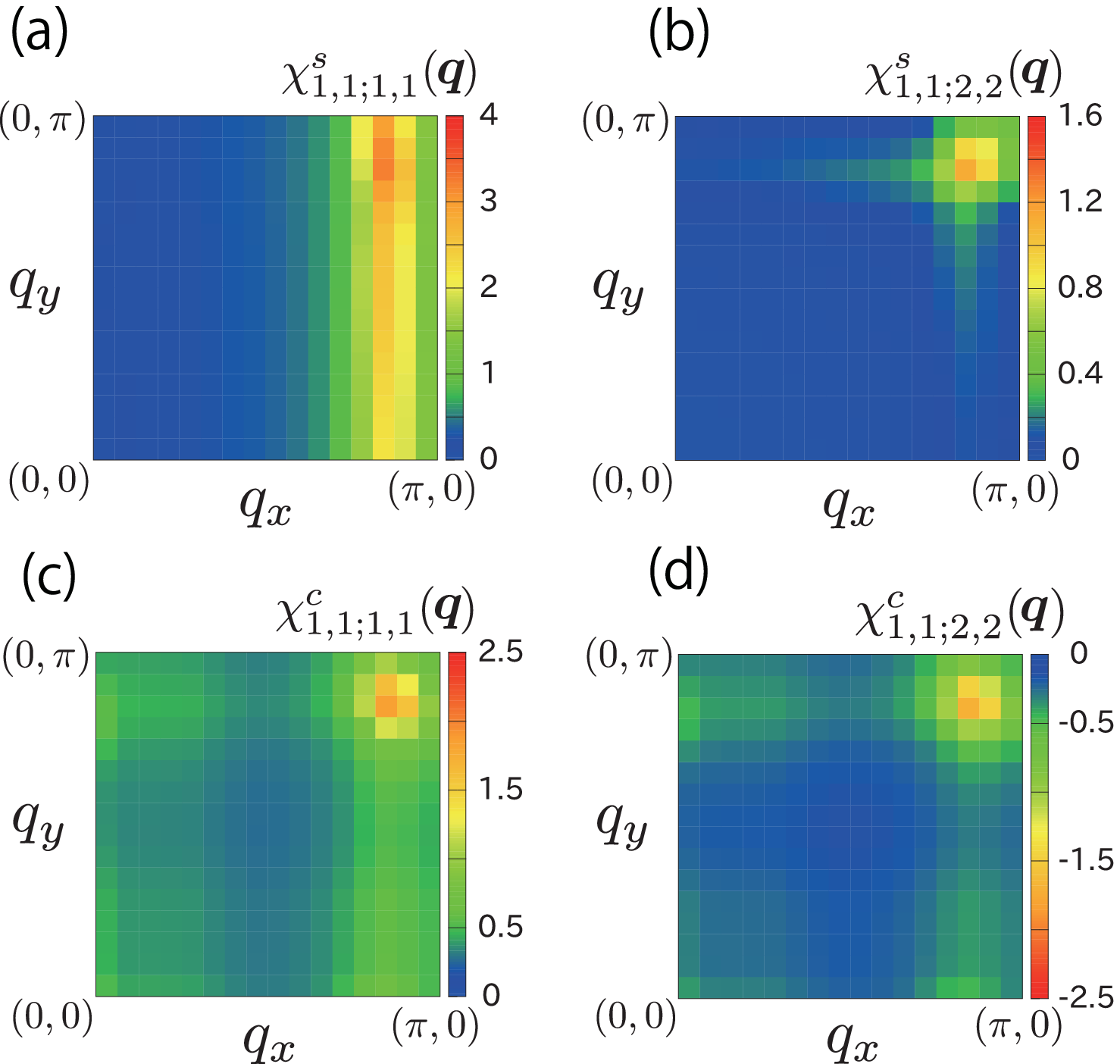}
\caption{
(Color online) 
The $\q$-dependence of the obtained susceptibilities
of (a) $\chi^{s}_{1,1;1,1}(\q)$,
(b) $\chi^{s}_{1,1;2,2}(\q)$, (c) $\chi^{c}_{1,1;1,1}(\q)$, and (d) $\chi^{c}_{1,1;2,2}(\q)$.
}
\label{fig:figsm0}
\end{figure}

Here, we explain the orbital dependence of the spin (charge) susceptibility
for $(U,g)=(2.1,0.15)$. The Stoner factors are $(\a_{S},\a_{C})=(0.92,0.93)$.
We show $\chi^{s(c)}_{1,1;1,1}(\q)$ and  
$\chi^{s(c)}_{1,1;2,2}(\q)$ in Figs. \ref{fig:figsm0}(a)-\ref{fig:figsm0}(d)
since $\chi^{s(c)}_{l,l';m,m'}(\q)$ has large value only for $l=l'$ and $m=m'$
in the present model.
We stress that $\chi^{c}_{x^2-y^2}(\q)$ with respect to the orbital polarization
$\Delta\hat{n}\equiv \hat{n}_{xz}-\hat{n}_{yz}$ is enlarged due to large negative value 
of $\chi^{c}_{1,1;2,2}(\q)$ at $\q\approx (0.8\pi,0.8\pi)$ shown in Fig. \ref{fig:figsm0}(d).
In contrast, the total charge susceptibility for the charge operator  $\hat{n}_{\rm tot}\equiv \hat{n}_{xz}+\hat{n}_{yz}$,
which is given as  $\chi^{c}_{\rm tot}(\q)=\sum_{l,m}\chi^{c}_{l,l;m,m}(\q)$, is not enhanced by the 
$\chi$-VC at all.

%

\begin{figure}[!htb]
\includegraphics[width=.98\linewidth]{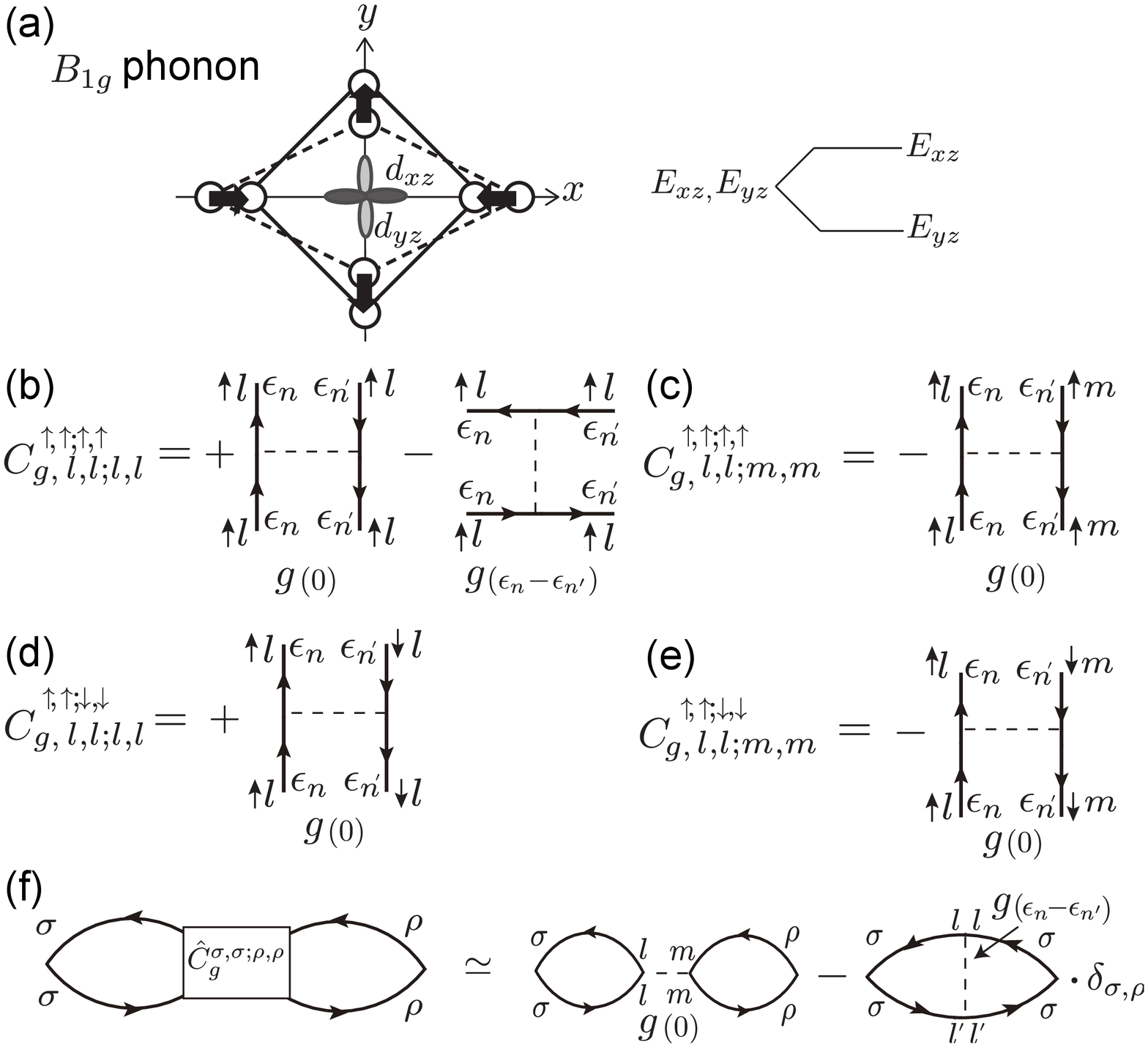}
\caption{
(Color online) 
(a) The electron-phonon coupling caused by $B_{1g}$-symmetry distortion.
The diagrammatic expressions for 
(b) $C_{g,\,l,l;l,l}^{\uparrow,\uparrow;\uparrow,\uparrow}$,
(c) $C_{g,\,l,l;m,m}^{\uparrow,\uparrow;\uparrow,\uparrow}$,
(d) $C_{g,\,l,l;l,l}^{\uparrow,\uparrow;\downarrow,\downarrow}$, and
(e) $C_{g,\,l,l;m,m}^{\uparrow,\uparrow;\downarrow,\downarrow}$.
Here, $l\neq m$.
(f) The first-order correction for $\hat{\chi}^{\s,\s;\rho,\rho}(q)$.
}
\label{fig:ladder}
\end{figure}

Next, we derive the matrix elements of the $B_{1g}$ phonon mediated interaction.
We show an example of schematic expression of $B_{1g}$ phonon mode in Fig. \ref{fig:ladder}(a).
In Figs. \ref{fig:ladder}(b)-\ref{fig:ladder}(e), we show the diagrammatic expression for the
spin-dependent phonon-mediated interaction term: 
$C^{\s,\s;\rho,\rho}_{g, \, l,l';m,m'}$.
Figure \ref{fig:ladder}(f) shows the first-order correction by $g$ for the spin-dependent susceptibilities $\hat{\chi}^{\s,\s;\rho,\rho}(q)$ at $\w_j=0$.
The first terms in Figs. \ref{fig:ladder}(b)-\ref{fig:ladder}(e) give the bubble diagrams, and the second term in Fig. \ref{fig:ladder}(b) gives the ladder diagram. 
In the case of $\omega_{D} \gg W_{\rm band}$, we can replace $g(\e_n-\e_{n'})$ with the constant $g$, so both the bubble and ladder diagrams contribute to the susceptibility.
Thus, the phonon-induced four-point vertex for $\omega_{D} \gg W_{\rm band}$ is
\begin{eqnarray} 
C^s_{g,\,l,l';m,m'}
&=&\begin{cases}
g&(l=l'=m=m')\\
-g&(l=m\neq l'=m')\\
0 &(\rm{otherwise})\end{cases},\label{eqn:Cs}\\
C^c_{g,\,l,l';m,m'}
&=&\begin{cases}
+g&(l=l'=m=m')\\
-2g&(l=l'\neq m=m')\\
+g&(l=m\neq l'=m')\\
0 &(\rm{otherwise})\end{cases},
\label{eqn:Cc}
\end{eqnarray}
where $\hat{C}^{c(s)}_{g}= \hat{C}^{\uparrow, \uparrow; \uparrow, \uparrow}_{g} + (-)\hat{C}^{\uparrow, \uparrow; \downarrow, \downarrow}_{g}$.
In the opposite case, $\omega_{D} \ll W_{\rm band}$, the ladder diagrams are expected to be small.
In fact, the corresponding irreducible susceptibility is approximately
$\chi_{l,l';m,m'}^{\rm ladder}(q) \approx -
\frac{T}{N_k}
\sum_{k} G_{l,m}(k+q)G_{m',l'}(k) \theta(\omega_{D}-|\epsilon_{n}|)$, which should be much smaller than $\chi_{l,l';m,m'}^{0}(q)$ due to $\theta(\omega_{D}-|\epsilon_{n}|)$.
For this reason, the phonon-induced four-point vertex 
for $\omega_{D} \ll W_{\rm band}$ is
\begin{eqnarray} 
C^s_{g,\,l,l';m,m'}
&=&0,
\label{eqn:Cs2} \\
C^c_{g,\,l,l';m,m'}
&=&\begin{cases}
+2g&(l=l'=m=m')\\
-2g&(l=l'\neq m=m')\\
0 &(\rm{otherwise})\end{cases}.
\label{eqn:Cc2}
\end{eqnarray}
Here, $\hat{C}^c_{g}$ in Eq. (\ref{eqn:Cc2}) corresponds to $ g_{l,l';m,m'}(\omega_{j})\equiv-2g(\omega_{j}) \cdot \delta_{l,l'} \delta_{m,m'}
(2 \delta_{l,m}-1)$ used in the main text. 
Since $\hat{C}^s_{g}=0$ in Eq. (\ref{eqn:Cs2}), $\a_{S}$ is independent of $g$ for $\omega_{D} \ll W_{\rm band}$ in the RPA.

In the SC-VC theory,
the charge-channel $\chi$-VC at $q=0$ is approximately proportional to 
$\sum_p \{ 3\chi^s(p)^2+\chi^c(p)^2\}$.
We have verified that the contribution to $\hat{\Lambda}^{c}$ from $\chi^s(q)$ dominates over that from $\chi^c(q)$ even for $\a_S\sim\a_C$.
For this reason, we can safely put $g=0$ in calculating the $\chi$-VC in the case of $\omega_{D} \ll W_{\rm band}$. 
%

\subsection{C: Expression for $U$-VC }


We explain the AL-type $U$-VCs,
which were also introduced in Ref. \cite{SRina}.
The charge- and spin-channel AL-terms in Fig. 1(d) in the main text 
are given as
\begin{widetext}
\begin{eqnarray}
	\Lambda^{{\rm AL}, c}_{l,l';m,m'} (k,k')
&=& \frac{T}{2N_{k}} \sum_{p} \sum_{a,b,c,d,e,f}
	G_{a,b} (k'-p) {\Lambda^0}'_{m,m';c,d;e,f} (k-k',p)
\nonumber \\
	& & \times
	\left\{
	    I^{c}_{l,a;c,d} (k-k'+p) I^{c}_{b,l';e,f} (-p) 
	 + 3 I^{s}_{l,a;c,d} (k-k'+p) I^{s}_{b,l';e,f} (-p)
	\right\}
\label{eqn:UALc} ,
\end{eqnarray}
\begin{eqnarray}
	\Lambda^{{\rm AL}, s}_{l,l';m,m'} (k,k')
	 &=& \frac{T}{2N_{k}} \sum_{p} \sum_{a,b,c,d,e,f} G_{a,b} (k'-p) 
	{\Lambda^0}'_{m,m';c,d;e,f} (k-k',p)
\nonumber \\
	& & \times
	\left\{
	  I^{c}_{l,a;c,d} (k-k'+p) I^{s}_{b,l';e,f} (-p)
	+ I^{s}_{l,a;c,d} (k-k'+p) I^{c}_{b,l';e,f} (-p)
	\right\}
\nonumber \\
	& & \quad
+ \delta\Lambda^{{\rm AL}, s}_{l,l';m,m'} (k,k')
\label{eqn:UALs} ,
\end{eqnarray}
where 
${\hat I}^{x}(q)= {\hat U}^{0;x}+{\hat U}^{0;x}{\hat \chi}^{x}(q){\hat U}^{0;x}$
in this SM.
The three-point vertex ${\hat \Lambda}^0(q,p)$ is given as
\begin{eqnarray}
&&\Lambda_{l,l';a,b;e,f}^0(q,p)
 =-\frac{T}{N_{k}}\sum_{k'}G_{l,a}(k'+q)G_{f,l'}(k')G_{b,e}(k'-p),
\end{eqnarray}
%
and 
${\Lambda^0}_{m,m';c,d;g,h}'(q,p)\equiv
\Lambda^0_{c,h;m,g;d,m'}(q,p)+\Lambda^0_{g,d;m,c;h,m'}(q,-p-q)$.
The last term in Eq. (\ref{eqn:UALs}) is given as
\begin{eqnarray}
	&&\delta \Lambda^{{\rm AL}, s}_{l,l';m,m'} (k,k')
	= \frac{T}{N_{k}} \sum_{p} \sum_{a,b,c,d,e,f} G_{a,b} (k'-p) 
	 I^{s}_{l,a;c,d} (k-k'+p) I^{s}_{b,l';e,f} (-p)
	{\Lambda^0}''_{m,m';c,d;e,f} (k-k',p) ,
\label{eqn:del-Lambda-ALs} 
\end{eqnarray}
where
${\Lambda^0}''_{m,m';c,d;g,h}(q,p)\equiv
\Lambda^0_{c,h;m,g;d,m'}(q,p)-\Lambda^0_{g,d;m,c;h,m'}(q,-p-q)$.
We verified that the contribution from
Eq. (\ref{eqn:del-Lambda-ALs}) is very small.

In addition, $\hat{X}^{x}(q)$
is written by using the $\hat{\Lambda}^{{\rm AL},x}(k,k')$ as follows
\begin{eqnarray}
X^{s(c)}_{l,l';m,m'}(q)=-\frac{T}{N_{k}}\sum_{k,a,b} G_{b,l'}(k)G_{l,a}(k+q)
 \Lambda_{b,a;m',m}^{{\rm AL},s(c)}(k,k+q)-\chi^{0}_{l,l';m,m'}(q).
\end{eqnarray}
\end{widetext}

Finally, we summarize the approximations applied to the numerical study in the main text.
In calculating the susceptibilities based on the SC-VC method, (i) we neglect the $\chi$-VC for spin channel $\hat{X}^{s}$, which has been justified in the five-orbital model as we discussed in Refs.
\cite{SYamakawa-eFeSe-SC,SYamakawa-FeSe} in detail. 
(ii) We also neglect the ladder diagrams due to the phonon-mediated interaction for the susceptibilities,
which is justified for $\omega_{D} \ll W_{\rm band}$.
In calculating the gap equation in the main text, (iii) we neglect the retardation effect by putting $g(\omega_{j})=g$ in the pairing interaction, and (iv) we drop the $U$-VC at finite $\omega_{j}$. The approximations (iii) and (iv) lead to the underestimation of the plain $s_{++}$ wave state, so the region of the full-gap $s_{++}$ wave state in Fig. 3(a) in the main text is underestimated.

\subsection{D: Local approximation of $U$-VC in the gap equation}

\begin{figure}[!htb]
\includegraphics[width=.7\linewidth]{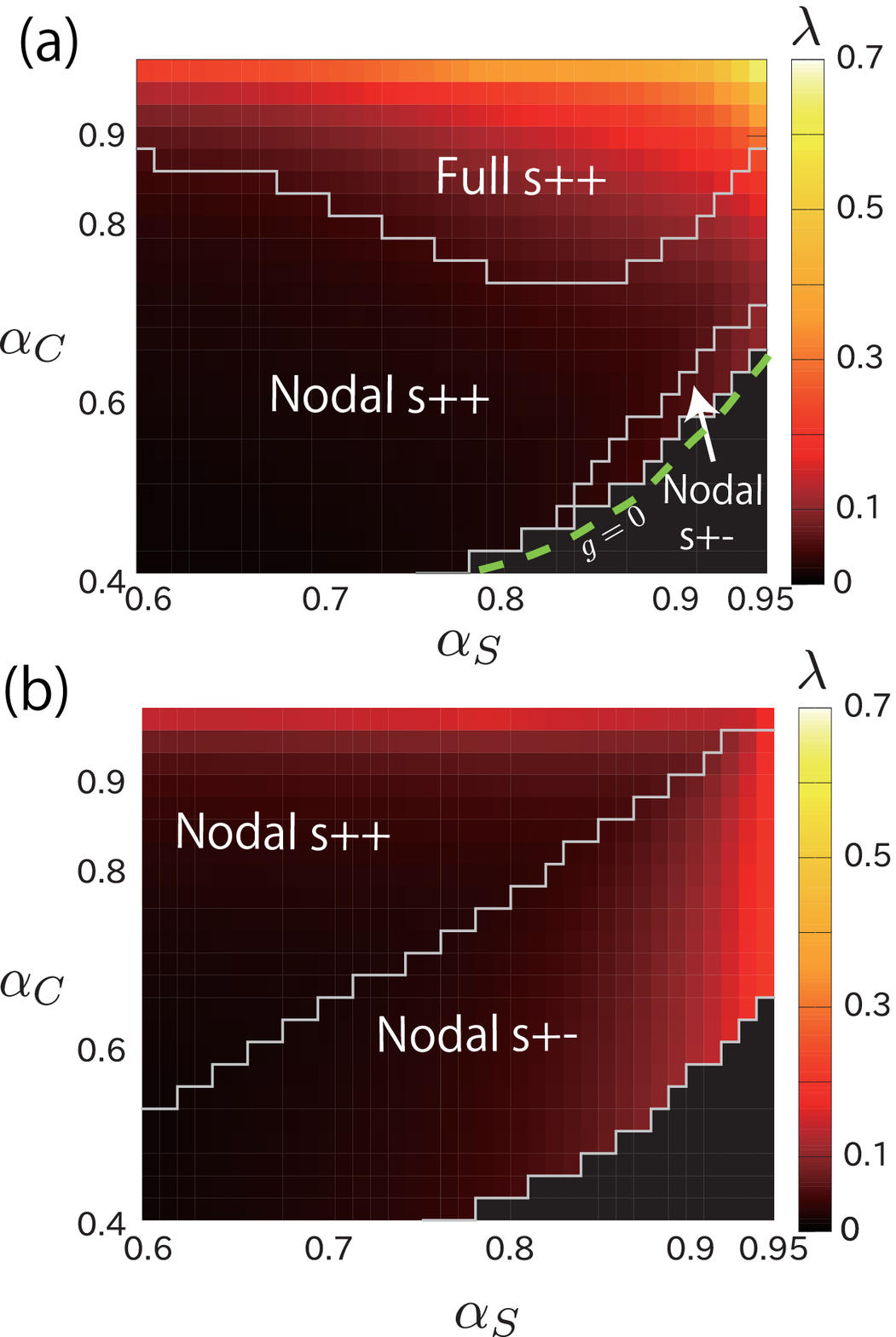}
\caption{
(Color online)
(a) The superconducting phase diagram in the $\a_S$-$\a_C$ space,
which is equivalent to the $\a_S$-$g$ phase diagram
in Fig. 3(a) in the main text.
(b) Obtained phase diagram by using the 
$U$-VC in the local approximation.
}
\label{fig:fig11}
\end{figure}

In the main text, we performed the numerical study of $U$-VC,
by taking account of its momentum dependence seriously.
Figure \ref{fig:fig11}(a) shows the superconducting phase diagram 
in the $\a_S$-$\a_C$ space, 
which is equivalent to the $\a_S$-$g$ phase diagram
in Fig. 3(a) in the main text.
However, this calculation is very time consuming, 
and it is very convenient  if the local approximation is applicable for the $U$-VC.
To check the validity of the local approximation,
we calculate the averaged $U$-VC over the FSs,
${\hat \Lambda}^x_{\rm loc}(\e_n,\e_{n'})=
\langle {\hat \Lambda}^x(k,k')\rangle_{\k,\k'\in{\rm FS}}$,
and analyze the gap equation by using this local $U$-VC.

Figure \ref{fig:fig11}(b) shows the obtained 
superconducting phase diagram by using the ${\hat \Lambda}^x_{\rm loc}(\e_n,\e_{n'})$.
We see that the full-gap $s_{++}$ state disappears
in this case, and this phase diagram is 
almost equivalent to that
given by the Migdal approximation in Fig. 3(e) in the main text.
Therefore, the momentum dependence of the $U$-VC has to be taken 
into account seriously in solving the gap equation.

\subsection{E: Filling dependence of the phase diagram}

\begin{figure}[!htb]
\includegraphics[width=.75\linewidth]{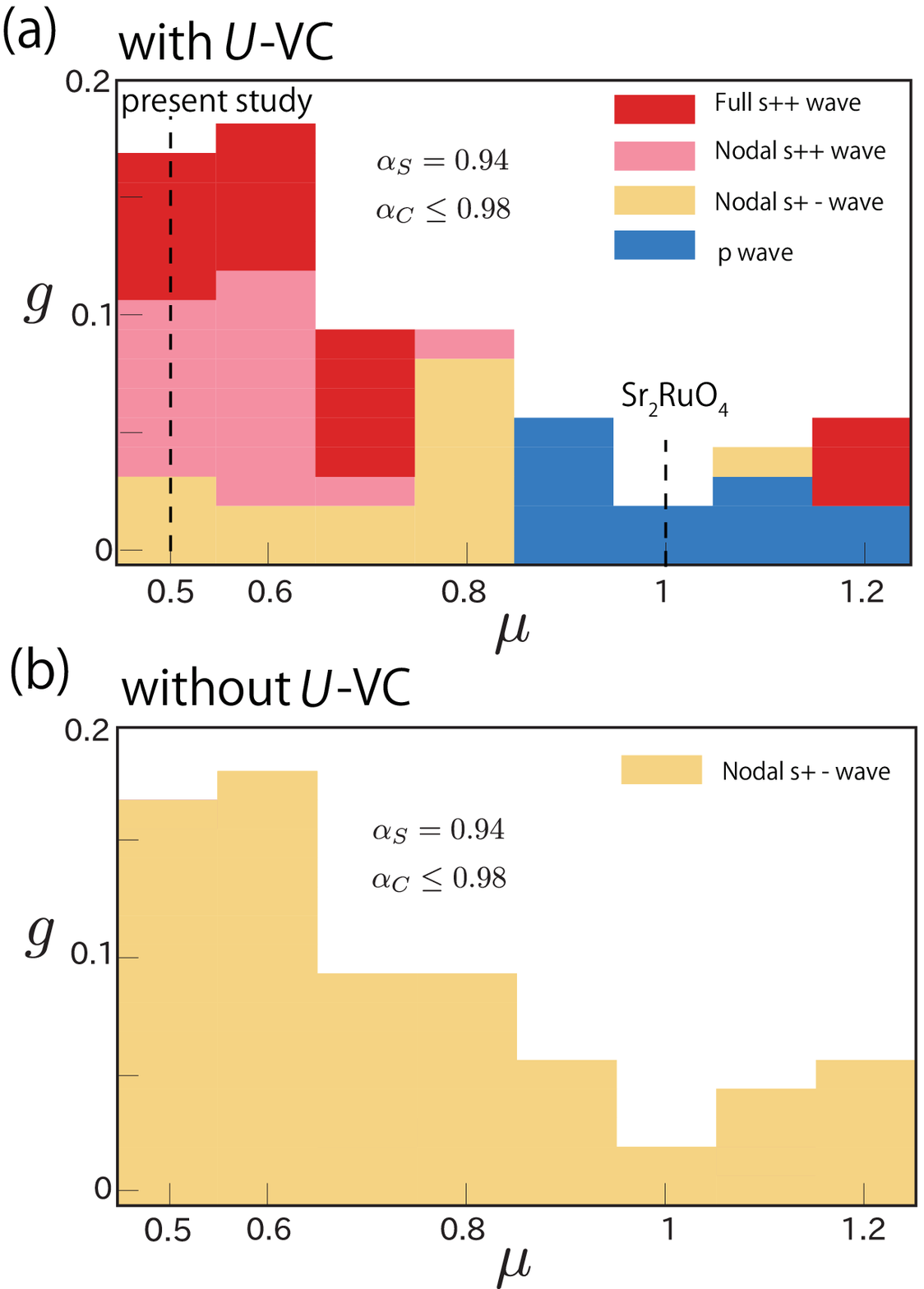}
\caption{
(Color online)
Phase diagram of the singlet and triplet states
 (a) with $U$-VC and (b) without $U$-VC
as functions of $\mu$ and $g$.
At each $\mu$, we set $U$ to satisfy the relation $\a_S=0.94$.
The electron filling for Sr$_2$RuO$_4$ ($n_{e}=2.67$ for FS $\a$, $\b$)
corresponds to $\mu=1.0$.
White color area corresponds to $\a_C > 0.98$.
The orbital order is realized for $\a_C > 1$.
}
\label{fig:fig4}
\end{figure}

In the main text, we show the superconducting phase diagram for the filling $n_{e}=2.30$.
Here, we show the filling dependence of the superconducting phase.
Figures \ref{fig:fig4}(a) and \ref{fig:fig4}(b) show the phase diagram of both
singlet and triplet states as functions of the chemical potential $\mu$ and $g$.
The self-energy is not included in the present study.
Here, $n_{e}=2.30$ corresponds to $\mu=0.50$.
At each $\mu$, we set $U$ to satisfy the relation $\a_S=0.94$.
The charge Stoner factor $\a_{C}$ increases with $g$, and the maximum value is set to $\a_{C}=0.98$.

In Fig. \ref{fig:fig4}(a), we show the obtained phase diagram when $U$-VC is taken into account.
We find that the present two-orbital model shows 
rich superconducting phase diagram, and the full-gap $s_{++}$ wave state
corresponds to the largest eigenvalue for a wide range of filling parameter.
It is noteworthy that the triplet superconductivity is appeared at $\mu \approx 1.0$, 
which corresponds to $\rm{Sr_2RuO_4}$.
This result is consistent with our previous study in Refs. \cite{SRina, STsuchiizu}.

On the other hand, the $s_{++}$ state disappears 
when we neglect the $U$-VC as shown in Fig. \ref{fig:fig4}(b). 
Thus, we conclude that the $U$-VC plays an important role in realizing the 
full-gap $s_{++}$ wave state for a wide parameter range.

\subsection{F: Retardation effect}\label{sec-a2}

In the main text, we studied the spin and orbital fluctuations
in the two-orbital Hubbard-Holstein model
with the phonon-mediated interaction $g(\w_j)=g\frac{\w_D^2}{\w_D^2+\w_j^2}$.
In solving the gap equation, we neglect the retardation effect.

\begin{figure}[!htb]
\includegraphics[width=.7\linewidth]{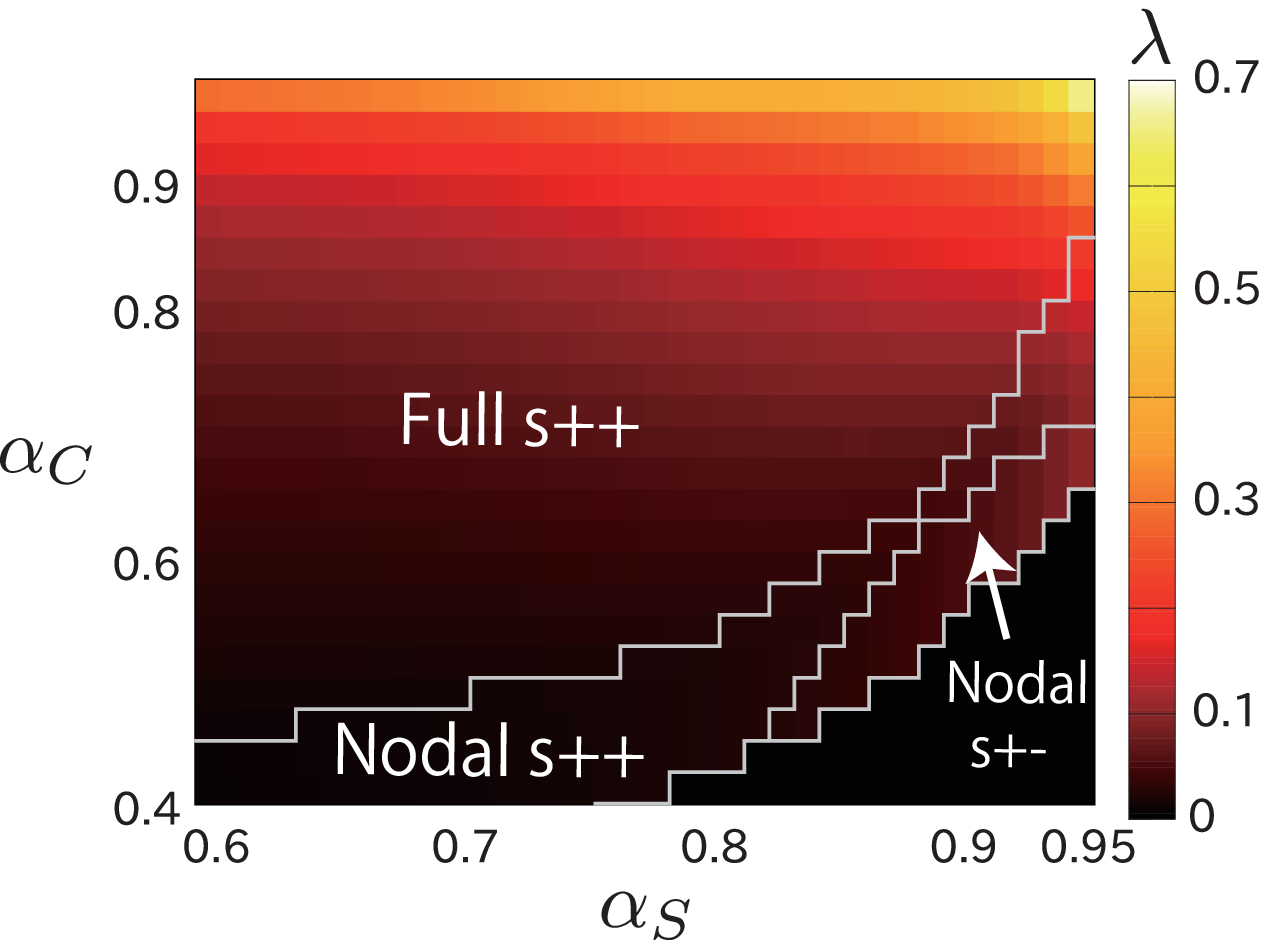}
\caption{
(Color online) 
The phase diagram obtained by taking account of the retardation effect.
The region of the full-gap $s_{++}$ state is expanded
by the retardation effect.
}
\label{fig:fig7}
\end{figure}

However, this simplification leads to the underestimation of 
the full-gap $s_{++}$ wave state.
Here, we study the case of $\w_D\ll T$, that is, $g(\w_j)=g\delta_{j,0}$.
In this case, the retardation effect becomes maximum.
The obtained phase diagram is given in Fig. \ref{fig:fig7}.
We find that the region of the full-gap $s_{++}$ state is drastically expanded
by the retardation effect.

\subsection{G: Impurity effect on superconductivity}

\begin{figure}[!htb]
\includegraphics[width=.7\linewidth]{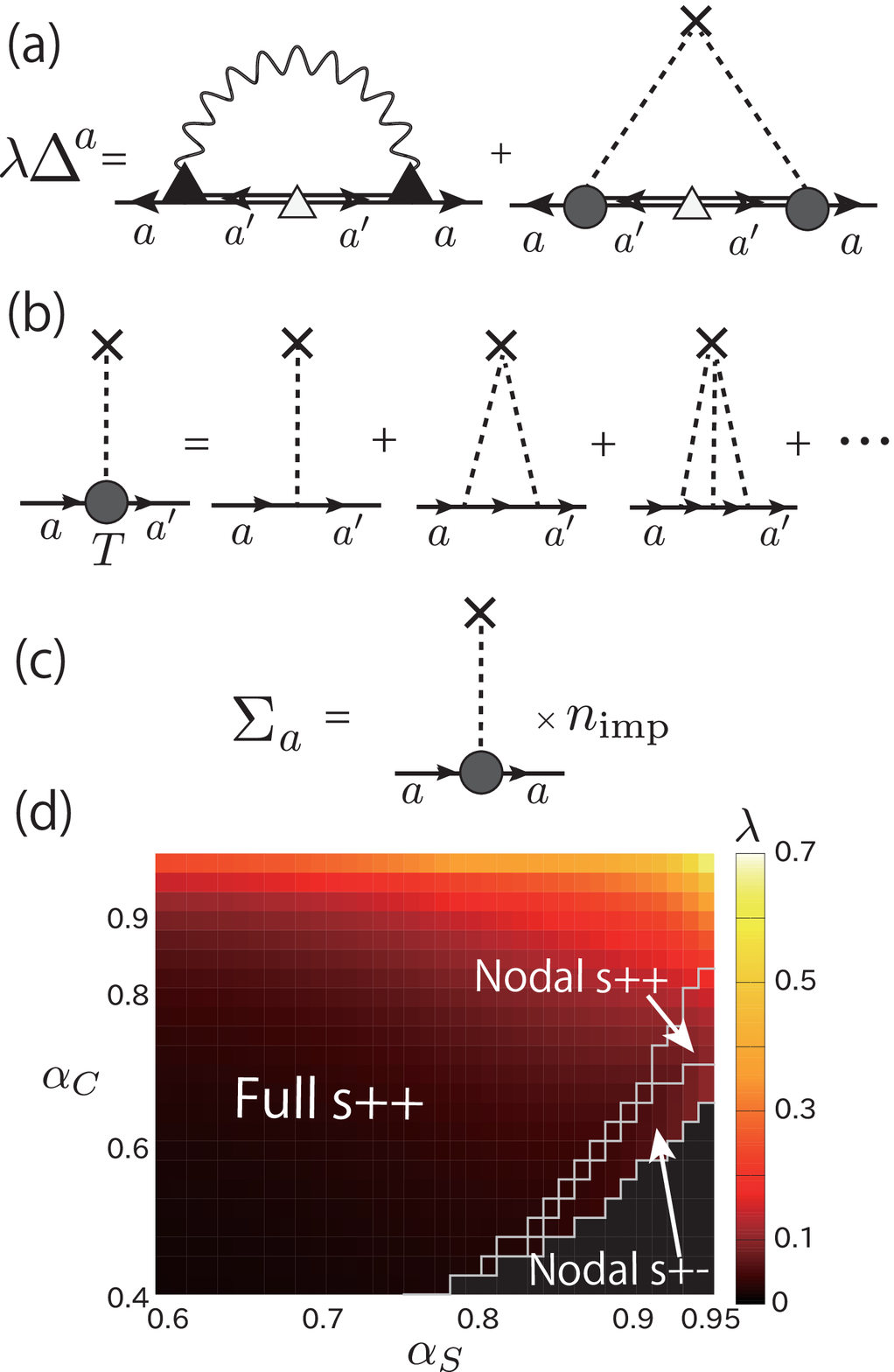}
\caption{
(Color online) 
(a) Gap equation in the presence of impurities.
(b) $T$-matrix given by the single impurity potential.
(c) Impurity-induced self-energy.
(d) The superconducting phase diagram for $n_{\rm imp}=0.1$\%
in the $\a_S$-$\a_C$ space.
The area of the full-gap $s_{++}$ wave state
is expanded by the impurity effect.
}
\label{fig:fig9}
\end{figure}

In the main text, we analyzed the superconducting gap equation 
based on the two-orbital Hubbard-Holstein model,
in the absence of the impurity effect.
However, it is well known that superconducting state
is sensitively affected by impurities.
Here, we analyze the gap equation in the presence of dilute 
non-magnetic impurities, using the $T$-matrix approximation.
The gap equation in the band basis is given as
\begin{eqnarray}
&&\lambda\Delta^{a}({\bm k},\epsilon_{n}) =- \frac{T}{N_k} \sum_{a',\epsilon_{m},\k'}
|G_{a'}({\bm k'},\epsilon_m)|^2 \Delta^{a'}({\bm k'},\epsilon_{m})
 \nonumber \\
&&\ \ \ \ 
\times \left[ {V}^{a, a'}_{s(t)}
({\bm k},\epsilon_n,{\bm k'},\epsilon_m) 
  -\frac{n_{\rm imp}}{T}|T_{a,a'}({\bm k}, {\bm k'}, \epsilon_{m})|^2 \delta_{n,m}\right],
\nonumber \\
\label{eqn:gap-eq-imp2}
\end{eqnarray}
which is schematically shown in Fig. \ref{fig:fig9}(a).
$V_{s(t)}$ represents the singlet (triplet) pairing channel,
which is given as $\hat{V}_{s}=3\hat{I}^{\Lambda.s}/2-\hat{I}^{\Lambda,c}/2$, and $\hat{V}_{t}=-\hat{I}^{\Lambda,s}/2-\hat{I}^{\Lambda,c}/2$. 
Here, $n_{\rm imp}$ is the impurity concentration,
and $T_{a,a'}({\bm k}, {\bm k'}, \epsilon_{m})$
is the impurity $T$-matrix with unitary scattering 
shown in Fig. \ref{fig:fig9}(b).
The Green function $G_{a}({\bm k},\epsilon_n)$,
expressed as a double line in Fig. \ref{fig:fig9}(a),
contains the impurity-induced normal self-energy on band $a$,
$\Sigma_a(k)=n_{\rm imp}T_{a,a}(\k,\k,\e_n)$,
shown in Fig. \ref{fig:fig9}(c).

Figure \ref{fig:fig9}(d) shows the superconducting phase diagram 
for $n_{\rm imp}=0.1$\% in the $\a_S$-$\a_C$ space.
We find that the area of the full-gap $s_{++}$ wave state
is drastically expanded by the impurity effect even for $n_{\rm imp}=0.1$\%.




\begin{thebibliography}{99}


\bibitem{Feng-eFeSe-swave}
C. H. P. Wen, H. C. Xu, C. Chen, Z. C. Huang, X. Lou, Y. J. Pu, Q. Song, B. P. Xie, Mahmoud Abdel-Hafiez, D. A. Chareev, A. N. Vasiliev, R. Peng, and D. L. Feng,
Nat. Commun. {\bf 7}, 10840 (2016).

\bibitem{Feng-eFeSe-swave2}
Y. J. Yan, W. H. Zhang, M. Q. Ren, X. Liu, X. F. Lu, N. Z. Wang, X. H. Niu, Q. Fan, J. Miao, R. Tao, B. P. Xie, X. H. Chen, T. Zhang, and D. L. Feng, 
Phys. Rev. B {\bf 94}, 134502 (2016).

\bibitem{A3C60}
Y. Takabayashi and L. Prassides, 
Phil. Trans. R. Soc. A {\bf 374}, 20150320 (2016).

\bibitem{Shen-replica}
S. Rebec, T. Jia, C. Zhang, M. Hashimoto, D. Lu, R. Moore, and Z. Shen,
Phys. Rev. Lett. {\bf 118}, 067002 (2017)

\bibitem{Millis}
Y. Zhou and A. J. Millis, Phys. Rev. B {\bf 93}, 224506 (2016).

\bibitem{Choi}
S. Choi, W.-J. Jang, H.-J. Lee, J. M. Ok, H. W. Choi, A. T. Lee, A. Akbari,
K. Nakatsukasa, Y. K. Semertzidis, Y. Bang, S. Johnston, J. S. Kim, and J. Lee, 
arXiv:1608.00886.

\bibitem{Jhonston}
L. Rademaker, Y. Wang, T. Berlijn, and S. Johnston, 
New J. Phys. {\bf 18}, 022001 (2016).

\bibitem{Capone}
M. Capone, M. Fabrizio, and E. Tosatti,
Phys. Rev. Lett. {\bf 86}, 5361 (2001).

\bibitem{A3C60-ph-theory1}
Y. Nomura, S. Sakai, M. Capone, and R. Arita,
 Science Advances {\bf 1}, e1500568 (2015).

\bibitem{Kim}
M. Kim, Y. Nomura, M. Ferrero, P. Seth, O. Parcollet, and A. Georges,
Phys. Rev. B {\bf 94}, 155152 (2016).

\bibitem{Matsuda-CeCuSi}
T. Yamashita, T. Takenaka, Y. Tokiwa, J. A. Wilcox, Y. Mizukami, D. Terazawa, Y. Kasahara, S. Kittaka, T. Sakakibara, M. Konczykowski, S. Seiro, H. S. Jeevan, C. Geibel, C. Putzke, T. Onishi, H. Ikeda, A. Carrington, T. Shibauchi, and Y. Matsuda,
arXiv:1703.02800.

\bibitem{Kittaka-CeCuSi}
S. Kittaka, Y. Aoki, Y. Shimura, T. Sakakibara, S. Seiro, C. Geibel, F. Steglich, Y. Tsutsumi, H. Ikeda, and K. Machida,
Phys. Rev. B {\bf 94}, 054514 (2016).

\bibitem{Onari-SCVC}
S. Onari and H. Kontani, 
Phys. Rev. Lett. {\bf 109}, 137001 (2012).

\bibitem{Kontani-RPA}
H. Kontani and S. Onari, 
Phys. Rev. Lett. {\bf 104}, 157001 (2010).

\bibitem{Takimoto}
T. Takimoto, Phys. Rev. B {\bf 62}, R14641(R) (2000). 

\bibitem{SM}
Supplementary Material

\bibitem{Yamakawa-FeSe}
Y. Yamakawa, S. Onari, and H. Kontani,
Phys. Rev. X {\bf 6}, 021032 (2016).

\bibitem{Yamakawa-eFeSe-SC}
Y. Yamakawa and H. Kontani, arXiv:1611.05375.

\bibitem{Rina}
R. Tazai, Y. Yamakawa, T. Tsuchiizu, and H. Kontani, Phys. Rev. B {\bf 94}, 115155 (2016).

\bibitem{Onari-SCVCS}
S. Onari, Y. Yamakawa, and H. Kontani, 
Phys. Rev. Lett. {\bf 112}, 187001 (2014).

\bibitem{Mig}
A. B. Migdal, J. Exptl. Thoret. Phys. {\bf 34}, 1438 (1958).

\bibitem{Ohno-SCVC}
Y. Ohno, M. Tsuchiizu, S. Onari, and H. Kontani, 
J. Phys. Soc. Jpn. {\bf 82}, 013707 (2012).
\bibitem{Pruschke}
O. Bodensiek, R. Zitko, M. Vojta, M. Jarrell, and T. Pruschke,
Phys. Rev. Lett. {\bf 110}, 146406 (2013).

\bibitem{Yamamoto}
K. Masuda and D. Yamamoto, Phys. Rev. B {\bf 91}, 104508 (2015).

\bibitem{Miyake}
K. Miyake and S. Watanabe,
arXiv:1704.00114.

\bibitem{Hattori}
K. Hattori,
J. Phys. Soc. Jpn. {\bf 79}, 114717 (2010).

\bibitem{Scalapino}
Z. B. Huang, W. Hanke, E. Arrigoni, and D. J. Scalapino,
Phys. Rev. B {\bf 68}, 220507(R) (2003).
\end{thebibliography}

\begin{thebibliography}{99}

\bibitem{SYamakawa-eFeSe-SC}
Y. Yamakawa and H. Kontani, arXiv:1611.05375.

\bibitem{SRina}
R. Tazai, Y. Yamakawa, T. Tsuchiizu, and H. Kontani, Phys. Rev. B {\bf 94}, 115155 (2016).

\bibitem{SYamakawa-FeSe}
Y. Yamakawa, S. Onari, and H. Kontani,
Phys. Rev. X {\bf 6}, 021032 (2016).

\bibitem{STsuchiizu}
M. Tsuchiizu, Y. Yamakawa, S. Onari, Y. Ohno, and H. Kontani, 
 Phys. Rev. B {\bf 91}, 155103 (2015).
\end{thebibliography}
\end{document}